\documentclass[]{spie}  

\usepackage{graphicx}
\usepackage{amsmath,amssymb}
\usepackage[usenames,dvipsnames]{color}
\usepackage[colorlinks=true, citecolor=blue, linkcolor=WildStrawberry]{hyperref}
\usepackage{epstopdf}

\newcommand{\beq}{\begin{equation}}
\newcommand{\eeq}{\end{equation}}
\newcommand{\bdm}{\begin{displaymath}}
\newcommand{\edm}{\end{displaymath}}

\newcommand {\smalltilde}{\raise.26ex\hbox{$\scriptstyle\mathtt{\sim}$}}
\newcommand {\micron}{$\mu \textrm{m}$}

\DeclareFontFamily{OT1}{pzc}{}
\DeclareFontShape{OT1}{pzc}{m}{it}{<-> s * [1.10] pzcmi7t}{}
\DeclareMathAlphabet{\mathpzc}{OT1}{pzc}{m}{it}

\graphicspath{{./plots/}}
\begin{document}

\title{A collimated beam projector for precise telescope calibration}

\author[1]{Michael Coughlin}
\author[2]{T. M. C. Abbott}
\author[1]{Kairn Brannon}
\author[3]{Chuck Claver}
\author[1]{Peter Doherty}
\author[4,5]{Merlin Fisher-Levine}
\author[3]{Patrick Ingraham}
\author[5]{Robert Lupton}
\author[1]{Nicholas Mondrik}
\author[1,6]{Christopher Stubbs}


\affil[1]{Department of Physics, Harvard University, Cambridge, MA 02138, USA}

\affil[2]{Cerro Tololo Inter-American Observatory, National Optical Astronomy Observatory, Casilla 603, La Serena, Chile}

\affil[3]{Large Synoptic Survey Telescope, Tucson, AZ 85718, USA}



\affil[4]{Brookhaven National Laboratory, Upton, NY 11973 USA}
\affil[5]{Department of Astrophysical Sciences, Princeton University, Princeton, NJ 08544, USA}




\affil[6]{Department of Astronomy, Harvard University, Cambridge MA 02138, USA}

\maketitle

\begin{abstract}

The precise determination of the instrumental response function versus wavelength is a central ingredient in contemporary photometric calibration strategies. This typically entails propagating narrowband illumination through the system pupil, and comparing the detected photon rate across the focal plane to the amount of incident light as measured by a calibrated photodiode. However, stray light effects and  reflections/ghosting (especially on the edges of filter passbands) in the optical train constitute a major source of systematic uncertainty when using a flat-field screen as the illumination source. A collimated beam projector that projects a mask onto the focal plane of the instrument can distinguish focusing light paths from stray and scattered light, allowing for a precise determination of instrumental throughput. This paper describes the conceptual design of such a system, outlines its merits, and presents results from a prototype system used with the Dark Energy Camera wide field imager on the 4-meter Blanco telescope. A calibration scheme that blends results from flat-field images with collimated beam projector data to obtain the equivalent of an illumination correction at high spectral and angular resolution is also presented. In addition to providing a precise system throughput calibration, by monitoring the evolution of the intensity and behaviour of the ghosts in the optical system, the collimated beam projector can be used to track the evolution of the filter transmission properties and various anti-reflective coatings in the optical system. 

\end{abstract}

\section{Introduction}
\label{sec:Intro}

Photometry is currently limited by precise broadband photometric calibration at the few percent level \cite{StTo2006}.
In particular, photometric calibration issues currently dominate the uncertainty budget for type Ia supernova cosmology \cite{BeKe2014,StBr2015}.
Photometric calibration of objects in images often still relies on observations of reference fields which lie outside the current field of view.
This makes traditional calibration techniques sensitive to systematic errors arising from the temporal and spatial variability of the atmosphere's optical transmission, as well as time-varying instrumental changes.
This is important for research areas that require sub-percent level photometric calibration such as exoplanet surveys, which search for small photometric perturbations, and cosmological supernova surveys, which require detailed knowledge of the relative instrumental response as a function of wavelength. 

The primary goal for the majority of calibration systems is to measure the dimensionless system transmission function $R_i(\lambda)$, where $\lambda$ is the wavelength of the light and $i$ is the pixel index. In general, this function is time-dependent and includes the effects of the atmosphere, optics, filter, and detector.
The time-dependent effects of the atmosphere can be directly measured by independent systems. The transmission properties of the atmosphere are determined by a combination of Rayleigh scattering in the atmosphere, small particle scattering from aerosols, and molecular absorption \cite{Houghton1977,Slater1980}. In addition, the atmosphere has significant narrowband emission at wavelengths longer than 750\,nm.
It is $R_i(\lambda)$ combined with the spectral photon distributions (SPDs) and the effective aperture of the pixels that determines the signal measured from a source.
The transmission function is affected by variations in internal interference giving rise to fringing, non-uniform interference filters, and water spots on detector anti-reflective coatings.
``Flat-fielding'' is traditionally used to normalize the variation in the pixels due to this function.
Because both the SPDs and the transmission function jointly determine the measured flux levels, there is no unique flat-field which can be used to normalize this variation, which is important for consistent analysis across visits.

Calibration is typically performed using flat-field screens, which are illuminated with light that fills the telescope aperture.
These over-fill the cone of angles which illuminates the telescope's focal plane.
Although flat-field screens are often exposed with lamps with sharp spectral features, one method \cite{StTo2006}, which is used by both the Pan-STARRS1 Survey \cite{StDo2010, ToSt2012}, SNLS \cite{ReGu2015}, and the Dark Energy Survey \cite{MoAr2012,MaRh2013}, is to use monochromatic light with a monitoring photodiode to determine each pixel's spectral aperture as a function of wavelength.
A limitation of this method, however, is that it does not account for ``scattered light,'' and its performance depends on the illumination pattern from the flat-field screen and the geometry of light scattering paths in the telescope and optics.
Spatial non-uniformity of the flat-field illumination can masquerade as photometric non-uniformity due to obscuration, dust on optics, degraded mirror coatings, or vignetting.
In addition, the scattered light from a flat-field screen has a scattering pattern which is very different from that from celestial objects.
There are therefore significant short-comings to relying on flat-field screens for accurate calibration, and thus alternate techniques are worth exploring.

One such technique is a collimated beam projector (CBP) which can be used to determine the instrumental response function and does not suffer from the same systematic scattered light effects.
In this design, a projector sends a beam of light with a diameter restricted by the size of the projector's aperture through the system. 
This is one component of the calibration system for the Large Synoptic Survey Telescope (LSST), which will rapidly and repeatedly survey one half of the sky with high-quality deep imaging and photometry \cite{Ivezic2014}. 
It has a three-mirror design with a f/1.2 beam, and will take rapid 15\,s exposures in optical photometric bands ($u$,$g$,$r$,$i$,$z$,$y$) similar to those used in the Sloan Digital Sky Survey (SDSS) \cite{FuYa2011}.
Photometric calibration is an essential component for achieving the survey's scientific goals, as LSST is likely to reach limits on systematic rather than statistical error in nearly all cases.
LSST has a number of particular advantages in achieving more accurate calibration.
The survey's rapid cadence allows for the use of celestial sources to monitor stability and uniformity of the photometric data, as each visit is exposed for significantly less time than the timescale on which the atmospheric extinction changes.
It has also been proposed that the Gaia mission could supply data to improve the photometric calibration of LSST \cite{AxMi2014}.
Spectroscopic measurements of atmospheric extinction and emission will be made continuously which will allow the broad-band optical flux observed in the instrument to be corrected to flux at the top of the atmosphere \cite{StHi2007}. 
On-sky photometric measurements will be used in conjunction with instrumental and atmospheric calibrations to calibrate the wavelength-dependence of the entire telescope and camera system throughput.

In this paper, we describe the design, implementation, and first tests of a collimated beam projector system.
The goals for the system are enumerated in section \ref{sec:goals}.
We describe the apparatus in section \ref{sec:apparatus}. 
Measurements and analysis are presented in section \ref{sec:results}. 
Section \ref{sec:Conclusion} concludes with a discussion of topics for further study.

\section{Goals}
\label{sec:goals}

The goal of the collimated beam projector is to determine the relative response of the telescope system with respect to wavelength over medium to large spatial scales. This is analagous to the determination of the ``illumination correction'' (low-order spatial sensitivity variations) that is often accomplished on-sky by rastering sources across the focal plane or through the {\"u}bercal process \cite{PaSc2008,ScFi2012}.
The collimated beam projector acts as an artificial star field, and enables the measurement of the relative system throughput in each filter, and allows the determination of instrumental zeropoints. 
This system can be used in conjunction with the more conventional flat-field screen approach, which is used to determine the high-spatial-frequency pixel-to-pixel quantum efficiency (QE) variations.
Furthermore, this device can be used to explore the effectiveness of the telescope system's baffling as well as to measure the ghosting in the optical train. 
The evolution of the ghosting intensity can be used to monitor any degradation in the quality of the AR coatings on the corrector optics and to measure any changes in filter transmission. 
The analysis goals of the collimated beam projector which drive the projector design are as follows:

{\em Ghosting.}
The collimated beam projector can distinguish between the direct and scattered light paths. Light is projected through one or more pinholes onto the telescope focal plane. Images are taken with a variety of filters with enough dynamic range so as to see both primary and secondary spots. Source extraction is performed on each image to determine the fluxes and positions for both the primary and secondary spots. Subsequently, the ratio of these spots is measured to determine the relative flux of the ghosts. The design requirement is that the spot size on the charge-coupled device (CCD) must be smaller than the closest ghost spot.

{\em Cross-talk.} 
The collimated beam projector can be used to allow calculation of the cross-talk coefficients for the CCDs in the system, which are non-zero due to the undesired coupling of signals owing to the proximity of their read-out electronics and cabling.
In general, this coupling occurs during the simultaneous read out of amplifiers, causing the imaged source in one amplifier to appear as a faint, mirror-symmetric ghost in another.
Images from light projected through multi-spot masks can be used to quantify this effect. 
By using a single filter with a variety of exposure times, we can measure measure cross-talk at various levels of CCD exposure. 
We can determine cross-talk in pixels (from baseline reset imperfections) by measuring the ratio of fluxes in nearby pixels.

{\em Throughput measurements.} 
A throughput measurement can be performed using a single pinhole projected at a single angle of illumination and placement on the filter, with the throughput measured by taking the ratio of the flux seen on the focal plane to the flux emanating from the beam projector. Measurements on other parts of the filter can be accomplished by repointing the projector. To characterize the filter's transmission alone, the projector can be re-pointed, keeping the illuminated area on the focal plane fixed, allowing the filter's throughput as a function position to be measured by taking the ratio of the fluxes seen on the focal plane, thus allowing the detector's sensitivity to cancel out. This measurement can be multiplexed by using a mask containing multiple pinholes which allows for illumination of multiple CCDs. A raster scan can be performed by repointing the projector. These measurements can be compared to flat-fields performed on the dome screen.

This single-ray illumination approach described has the disadvantage that the full pupil is not illuminated. This, in turn, means that not all angles will be illuminated; however, it is possible to take a set of data that probes the full phase space.
Taking, for example, the size of the collimated beam that illuminates a single pixel to be 180\,mm, we would require a total of 
$N = \frac{\textrm{LSST effective aperture}}{\textrm{Beam Area}} = \left(\frac{6.5\,m}{0.18\,m}\right)^2 = 1300$\,``pointings.'' At a nominal imaging cadence of 20 seconds per frame, this would take about 7 hours (or about one cloudy night of time). 
In addition, the full scan described above is unlikely to be necessary. To the extent that the system is axisymmetric, a radial scan which probes the entire illumination cone through the optical train would suffice.
This radial scan would only require $N = \left(\frac{3\,m}{0.18\,m}\right) = 17$\, pointings. The radial scans would then be weighted in proportion to the effective collection area at each respective radius.

{\em Sensor characterization.} 
Sensor effects such as tree rings, the ``brighter-fatter'' effect, and other effects resulting in astrometric and photometric perturbations can also be measured \cite{Rasmussen2014,Stubbs2014,Walter2015}.

{\em Rotation about collimated beam projector pupil.}
In order to use the multi-aperture masks to tie together the response across different parts of the focal plane, we need to know the relative flux being broadcast by each pinhole. We have arranged the most recent version of the prototype collimated beam projector to allow us to rotate the projector about the pupil of the collimating optic. This allows us to issue the beam from each pinhole through the same path in the optical train of the main telescope, placing each spot in turn on the same portion of the primary mirror and on the same location of the focal plane. This therefore allows us to measure (using the integrating sphere's monitoring photo-diode to ensure flux stability) each spot's intensity relative to the others. We achieved this by constructing a custom alt-az mount with the intersection of the rotation axes placed at the lens pupil. 

{\em Melding dome flats with collimated beam projector data.}
The information provided by monochromatic dome flats and the collimated beam projector images are, in many respects, complementary. The dome flats provide full-field illumination over the entire input pupil, but are contaminated by illumination from non-focusing light paths. The collimated beam projector spots are free from stray light, but only achieve partial pupil-illumination at discrete locations. Moreover, even without stray light problems, using dome flats to establish flux calibration across the focal plane is limited by the uniformity of the integrated surface brightness over different regions of the screen, since it does not reside at the input pupil of the system. 

We plan to use a combination of the collimated beam projector spot images and monochromatic dome flats to exploit the desirable attributes of each data set. At a given wavelength, we will generate a set of collimated beam projector images that appropriately sample the input pupil of the system, by rastering the portion of the primary mirror that is illuminated by the collimated beam projector beam, while adjusting the orientation of the collimated beam projector and the telescope so as to keep the spot locations fixed on the focal plane. This will ensure, for example, that we appropriately sample the cone of rays incident on the interference filters. This pupil-averaged collimated beam projector spot image will then be used to produce a ghost-corrected monochromatic dome flat. In effect, we will use the dome flats to interpolate the system response function measurements for regions of the focal plane that lie between the collimated beam projector spots. This amounts to making an in-dome illumination correction using virtual stars from the collimated beam projector. Of course, one could supplement this using actual stars on the sky, but the advantage of this approach is that we can use the in-dome systems to do this in narrowband light, whereas rastered sky images can only be obtained in the relatively broad survey passbands, combined with the object's intrinsic ``spectral energy distribution.''

\section{Apparatus}
\label{sec:apparatus}

\begin{figure}[t]
\centering
 \includegraphics[width=5.0in,angle =0]{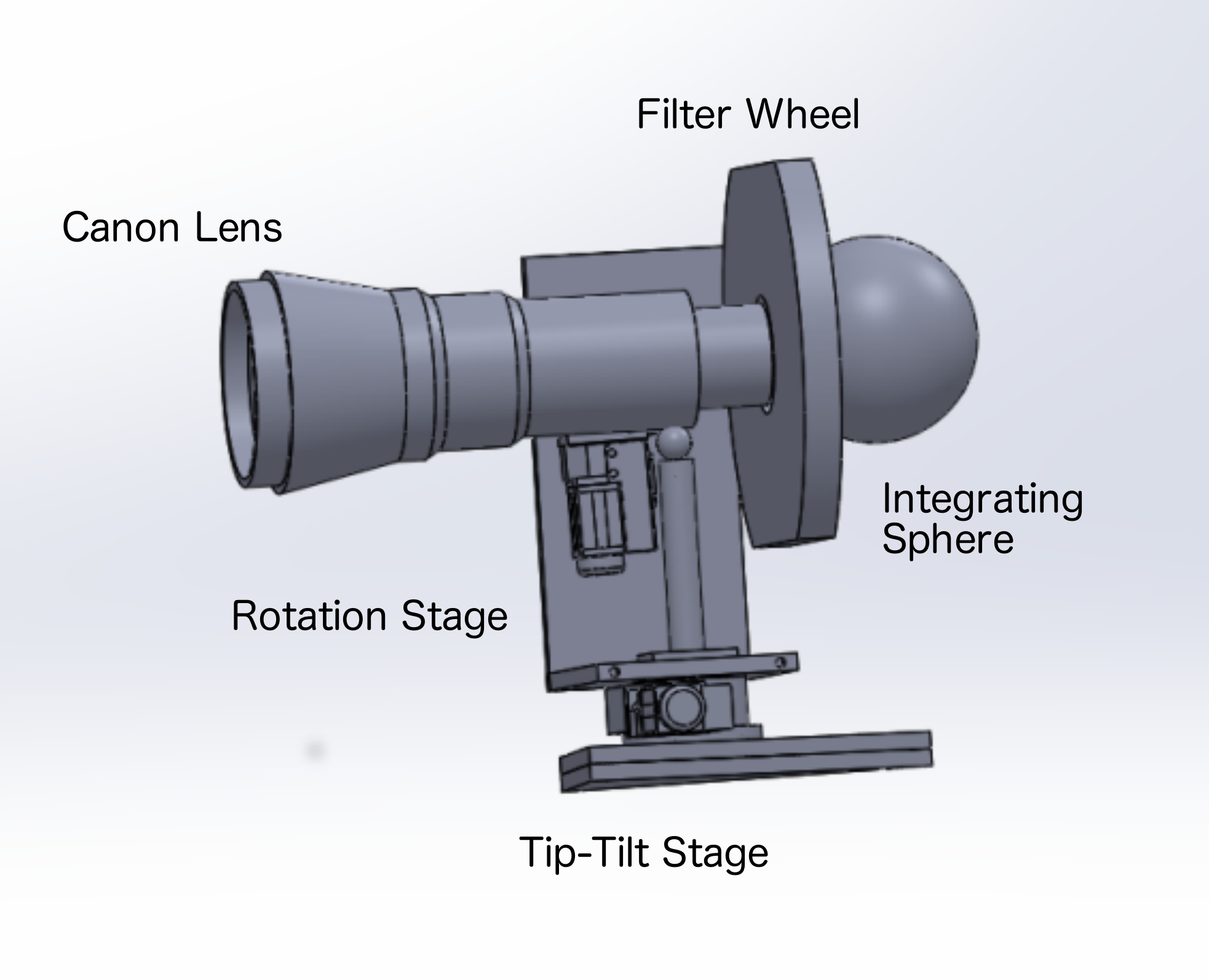}
 \caption{Sketch of the apparatus. The apparatus is mounted on a alt-az rotation stage and includes a light source, an integrating sphere, a mask/filter wheel, and the collimating optic. For the sake of scale, the lens diameter is 5\,in.}%
 \label{fig:apparatus}
\end{figure}

The instrument projects a collimated beam of light of known shape, wavelength, and intensity onto the telescope focal plane.
Figure~\ref{fig:apparatus} shows the basic design concept.
An optical fiber brings light (monochromatic or any other desired spectral distribution) from a source through a shutter to an integrating sphere. 
This homogenizes any wavelength-dependence in the angular distribution from the source. 
The light then travels to a filter wheel containing a set of ten filters/masks which allows for a choice between a variety of spatial patterns. 
After the light has been modified it travels through the collimating telescope and out of the system. 
The light is then re-imaged onto the telescope focal plane. 
The collimator optic will reside inside the enclosure on an alt-az mount so that the collimated beam can be re-imaged onto any location on the focal plane. With the additional freedom of moving the telescope, the collimated beam can achieve any desired position within the input pupil. Consequently, the entire 4-d phase space of input ray positions and angles can, in principle, be scanned.

Our prototype collimated beam projector system has undergone significant evolution, with three phases. The first was used for the calibration of the PanSTARRS system, with results described in Stubbs et al. \cite{StDo2010} and Tonry et al. \cite{ToSt2012}. The second iteration used a Takahashi ED-180 astrograph as the collimating optic, with an f/2.8 beam and a 500 mm focal length, placed on a commercial Paramount MX+ mount with a Thorlabs PDA200C pre-amplifier used to monitor the CBP integrating sphere photocurrent. The most recent iteration uses a Canon EF 500mm f/4 lens as the collimating optic, on a custom alt-az mount which pivots about the lens pupil. We also improved the precision of the photodiode monitoring electronics, with a Keithley 6514 Programmable Electrometer used to monitor the collimated beam projector integrating sphere photocurrent. Lastly, the most recent version uses a monochromator (Newport part number 74125) as the light source, rather than a white light source and narrowband filters. The experience gained with these systems has informed the design of the collimated beam projector that will be used for the LSST project, and has also produced a portable prototype system that we are using to characterize various existing telescope/instrument combinations.  

We now describe each component in turn, following the light path.
The light source used with the collimated beam projector is not restricted; one could use a white light source, tunable laser, monochromator, or laser diode.
In the second version, we use a white light source with both narrowband (10\,nm) and broadband Sloan Digital Sky Survey filters (to approximate the LSST filters).
In the third version, we use a monochromator as the light source.
A shutter is placed on the output ports of the light sources.
The most important design consideration for the light source is that it results in a sufficient surface brightness projecting onto the telescope's CCD focal plane.

The light then leaves the light source by way of broadband (340-800\,nm and 420-2000\,nm) Newport light-guides, to direct the light into the integrating sphere.
The integrating sphere ensures that the aperture mask has uniform surface brightness and that the calibration light retains no history from the upstream illumination system.
It contains a photodiode to monitor the output light, which determines the flux from the source onto the telescope focal plane. 
This can be used to correct for variations in the light intensity masquerading as variations in the throughput.
As calibrated photodiodes with National Institute of Standards and Technology (NIST)-traceable metrology standards with spectral response known at the 0.1\% level over the wavelengths relevant for CCD instruments exist, it is possible to map the sensitivity of the apparatus as a function of wavelength at this level. 

\begin{table}[htp]
\begin{center}
\begin{tabular}{|c|c|c|}
\hline
Filter Wheel Slot 	& Filter	& Mask \\
\hline
1 & 568\,nm & 200$\mu m$ slit \\
2 & 700\,nm & 20$\mu m$ pinhole \\
3 & 671\,nm & Ronchi grating \\
4 & 2.2$^\circ$ field stop & 20$\mu m$ multi-pinhole \\
5 & 680\,nm & USAF target \\
\hline
\end{tabular}
\end{center}
\caption{Masks and filters available for the collimated beam projector. The narrowband filters have 10\,nm widths. }
\label{tab:masks}
\end{table}%

Light passes from the integrating sphere to a filter wheel, which can accommodate ten 50 mm $\times$ 50 mm filters or masks.
It will be useful to have the ability to illuminate the entire 3.5 degree diameter LSST field, and this corresponds to a 30.55 mm $\times$ 30.55 mm square aperture at the focus of the f/2.8 collimated beam projector. 
The aperture mask in the filter wheel determines the shape of the light distribution on the focal plane of the telescope.

In the analysis that follows, we will use single and multi-precision pinhole arrays of 20$\mu m$ circles.
The available masks and filters in the following analysis are described in Table~\ref{tab:masks}.
The physical size of the mask holes are determined by a balance of needing enough light to emerge from a single pinhole to measure with high signal-to-noise ratio on the focal, plane and not having spots so large as to have difficulty discriminating between the direct light path and the ghosted/scattered light paths. 
The aperture mask at the focus of the projector is re-imaged onto the instrument with a magnification that is given by the ratio of the focal lengths.
Both the first and second collimated beam projector optical systems used for this demonstration had a 500\,mm focal length, and therefore the Blanco 4\,m, with a 10\,m focal length, has a magnification factor of 20$\times$. Similarly, the NOFS 1.3\,m with a 5.2\,m focal length has a magnification factor of 10.4$\times$.


The use of an aperture mask consisting of an array of pinholes allows for multiplexing throughput measurements. 
We elect to place a spot onto the center of each detector in the focal plane because differences in silicon lots and between device AR coatings can produce chip-dependent changes in QE. 
This corresponds to a rectilinear array of pinholes on a grid with spacing corresponding to the spacing between the detectors on the focal plane.
The LSST CCD detectors consist of \smalltilde 4000$\times$4000 10\micron\, pixels. 
Placing a spot on each detector would then require a 15$\times$15 array of pinholes with a spacing of 40\,mm / 20.4 = 1.96\,mm. 
Repointing the collimated beam detector can shift this pattern on each detector.

After the light passes through the mask, the light continues through the collimating optical system. We have used two such collimators, the first a commercial Takahashi wide field astrograph, which is a f/2.8 hyperbolic Newtonian system (plus corrector) that provides excellent optical quality over a 5 degree flat focal plane.
The collimator has a focal length of 0.5\,m, a diameter of 180\,mm, and a 5\,deg FOV diameter. 
The second is a Canon EF 500mm f/4 lens.
This compares to the LSST specifications of a 10.2\,m focal length, a plate scale of 0.02 arcsec per \micron, f/\# of 1.2, and a 3.5\,deg FOV diameter.

The procedure for performing measurements is as follows.
To align the collimated beam projector, the telescope is first pointed directly at the platform on which the projector is mounted.
A digital level is then used to point the projector at the same altitude as the telescope.
Then, with the mirror covers closed, a laser pointer is used to determine where the light from the projector will fall on the primary mirror.
Afterwards, a Ronchi grating is placed in the mask position, which allows a significant fraction of the light through and contains structure that can be seen by eye in the images.
At a light level significantly below the camera's full-well, the collimated beam projector is rotated in azimuth until the pattern can be identified in the telescope images. 
At this point, the mask is changed to the multi-pinhole array, which, as described above, is designed to place a spot on each CCD.
This mask was designed to maximize the number of resolvable spots without creating overlapping ghosts in any given frame, which makes it difficult to disentangle their effects.
Small adjustments in both altitude and azimuth, in addition to rotations of the mask, are performed iteratively in order to center as many spots as possible.

\section{Demonstration}
\label{sec:results}

\subsection{Collimated Beam Projector calibration}


\begin{figure}[t]
\centering
 \includegraphics[width=5.0in,angle =0]{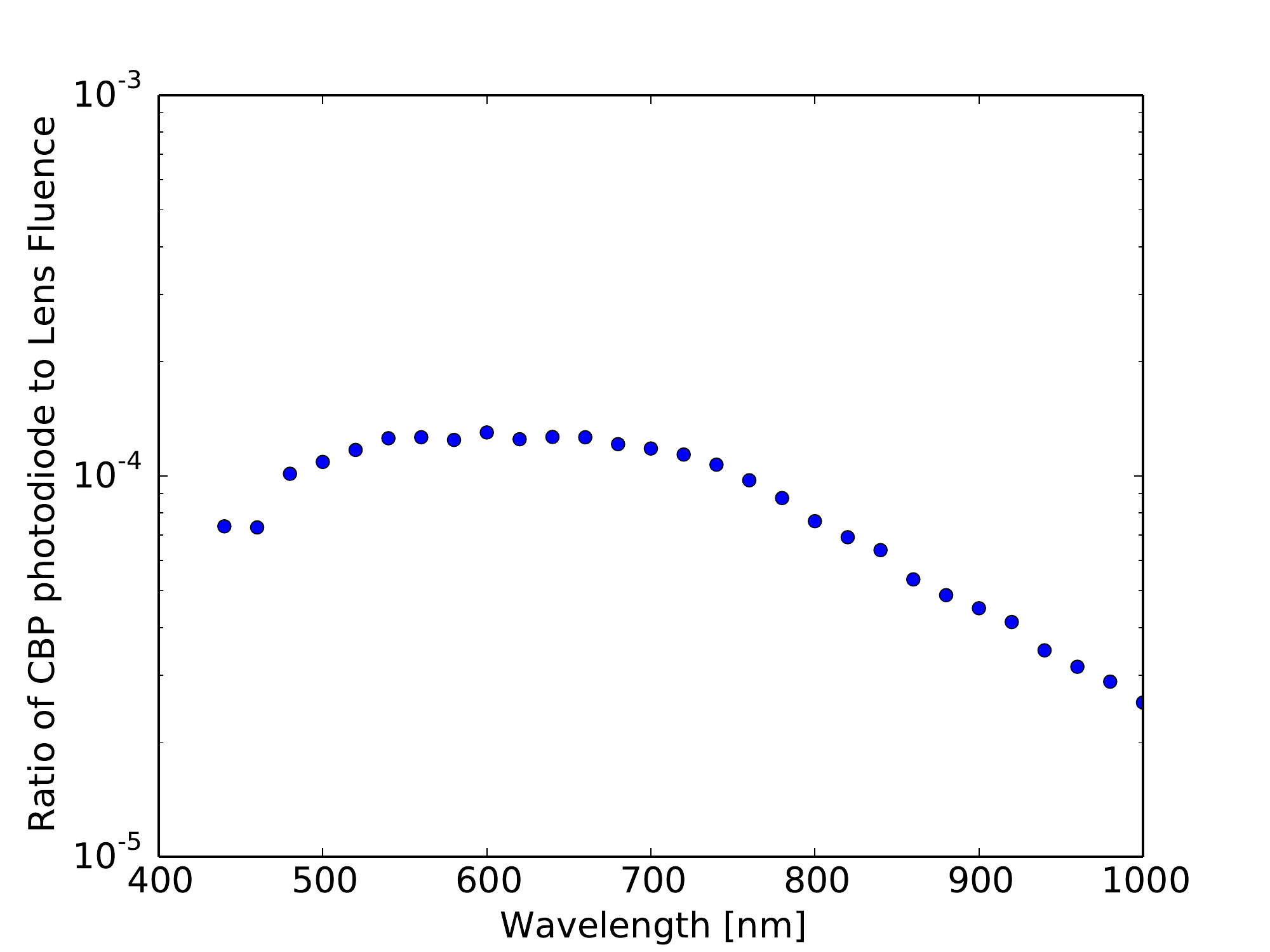}
 \caption{Ratio between the collimated beam projector integrating sphere monitoring photodiode flux and the output fluence from the system. This allows for the mapping of the collimated beam projector monitoring photodiode to the number of photons emerging from the system, measured with a NIST-calibrated photodiode.}%
 \label{fig:cbpthroughput}
\end{figure}

The first goal is to measure the relative throughput of the collimated beam projector system. 
In this analysis, we use the third version of the collimated beam projector.
The monochromator is connected to a shutter and light-pipe entering the integrating sphere.
The light in the integrating sphere of the collimated beam projector is measured with a photodiode installed in the integrating sphere.
This gives a measure of the amount of light entering the system.
The light then passes through the optical system. 
It is projected into another (larger) integrating sphere which captures the entire beam, and is equipped with a second monitoring NIST-calibrated photodiode.
We measure the (dark-current subtracted) photocurrent from both photodiodes.
At a single wavelength, we measure the dilution factor from measuring the output at an integrating sphere port.
Combining these measurements, we determine the fluence emerging from the collimated beam projector system, which can be monitored by the photodiode installed in the integrating sphere.
Figure~\ref{fig:cbpthroughput} shows the system flux calibration, which is calculated from the ratio of these currents.

\subsection{Dark Energy Camera Measurements}

\begin{figure}[t]
\centering
 \includegraphics[width=5in]{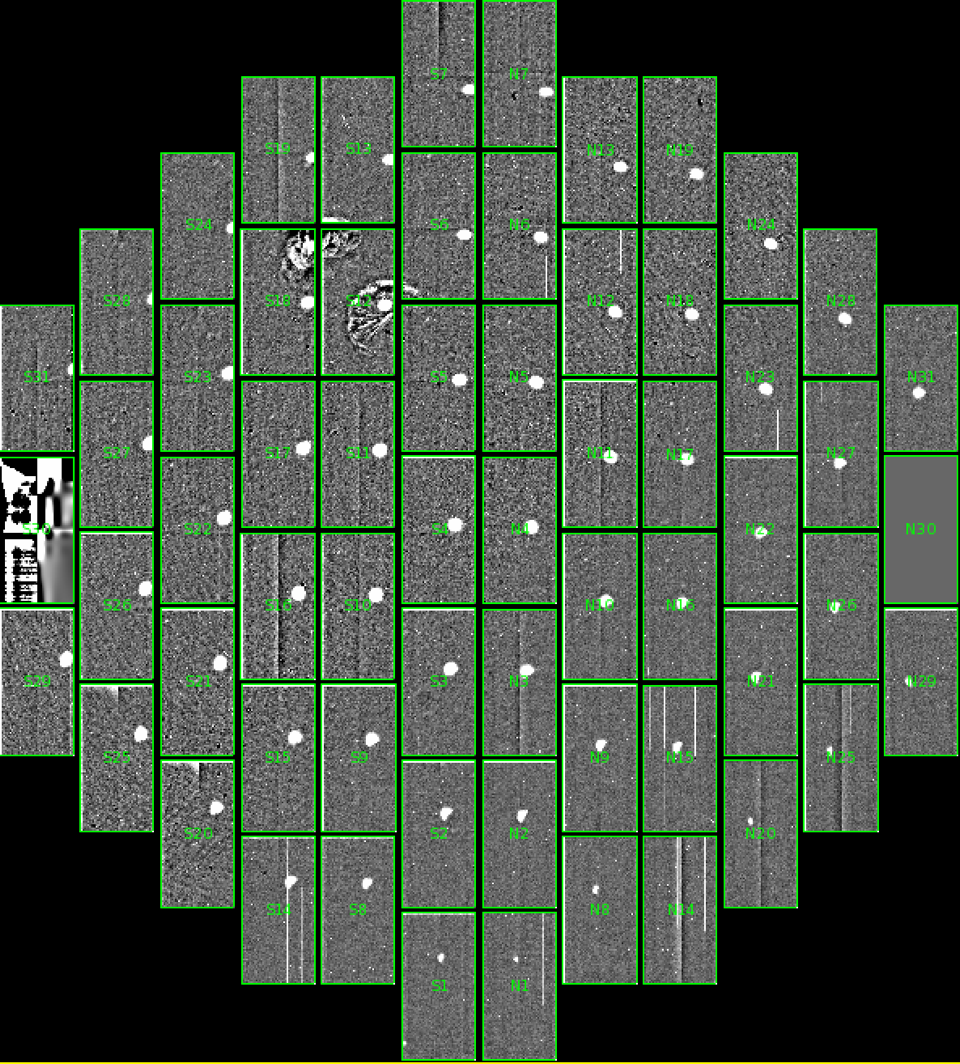}
 \caption{An example $r$-band image of the Dark Energy Camera focal plane with each of the CCDs illuminated with a spot from the CBP multi-pinhole mask.
}%
 \label{fig:DECam}
\end{figure}	

Figure~\ref{fig:DECam} shows an example of a Dark Energy Camera image with a collimated beam projector multi-pinhole array, using the second version of the projector.
The multi-pinhole images show a number of interesting features.
Each pinhole appears as a bright source on an individual CCD, and these spots have several novel applications.

\subsubsection{Optical and cross-talk ghosting}
Optical ghosts and cross-talk ghosts, though arising from fundamentally different origins, can result in similar effects, which are often hard to disambiguate. Cross-talk ghosts always appear at fixed pixel-coordinate offsets with respect to their aggressor, due to their production by couplings in the readout electronics, whereas optical ghosts and glints arise from multiple reflections in the optical system, and therefore move with respect to their aggressor as a function of the location of the aggressor on the focal plane (or, more accurately, their path through the optical train). The end result, however, is a faint, secondary image of a bright star at some unknown offset.

The collimated beam projector plays a critical role in the characterization of the cross-talk and optical ghosting in the system, as well as allowing disambiguation of these two effects. To measure cross-talk ghosts, and thus determine the coupling coefficients of the cross-talk matrix used for their removal, a sparse array of small, bright spots is dithered around the focal plane, and the faint ghosts which arise are measured. Performing these measurements on-sky using stars involves disentangling real sources from fake ones, and this is hard, if not impossible, to do. The collimated beam projector allows for the placement of an arbitrarily sparse array; a single spot can be moved around the focal plane, and the cross-talk ghosts which arise can be measured. This spot can then be kept in a fixed position on the focal plane, whilst changing its path through the optical train. Ghosts which don't move are known to arise from cross-talk, whereas ghosts which do move are known to be optical in nature. In practice, most of the coupling matrix elements will be small, and so the measurement can be sped up with multiplexing, \emph{i.e.} putting one or more spots down on each CCD simultaneously, as illustrated in Figure~\ref{fig:DECam}.

\subsubsection{Throughput measurements}

The collimated beam projector is also used to measure the throughput of the system, with the analysis proceeding as follows. After overscan subtraction, we fit the local background around each spot, computed from the pixels in the CCD which contains the spot in question. Due to the irregularity of the holes in the mask and the imperfect alignment of the mirrors within the collimated beam projector, PSF fluxes cannot be used, and the flux is calculated by simply summing the pixels over threshold in the spot. We note that while this procedure works well for the bright spots, it can be problematic for measuring the ghosts which have significantly less flux, as peak signal levels can be barely above background despite being visible to the eye. We have verified by eye that the automated centroiding and flux measurement gives sensible results, but this tends to fail for ghosts, which usually require special treatment ``by hand.''

The goal is a relative throughput measurement in the broadband filters, sampled at each of the narrowband filter passbands available. 
This is a proxy for when a monochromator will be available to measure filter transmission in small steps across the passband.
For a given pointing in a DECam broadband filter (taking the $r$-band filter here as an example), images are taken in each narrowband filter, and the flux measurements from each spot in the different passbands are compared. The flux values in the narrowband filters differ by factors of 42, 12, 15 and 14 for the 568\,nm, 671\,nm, 680\,nm and 700\,nm filters respectively when compared to the flux with no narrowband filter in the CBP. The three closely spaced filters sit in the flattop region of the DECam broadband filter, and therefore all have similar throughput, whereas the 568\,nm filter sits on the band-edge, leading to much lower transmission.

Due to the ability to tune the spot brightness on the focal plane, the statistical error on measurements can be made relatively small.
Therefore, understanding the systematic uncertainty in these measurements is important.
One source of systematic error arises from the potential for out-of-band light leak across the different filter bands.
This is due to a combination of quantum efficiency variation across a filter (i.e. efficiency near 100\% in the passband and near 0.1\% out-of-band) and variation in flux with respect to wavelength from the light source, as the one used emitted more at the red wavelengths than at blue wavelengths.
Due to the fact that we used narrow-band filters (10\,nm FWHM) at 568, 671, 680, and 700\,nm, the variation due to red leak, which is due to light from wavelengths longer than the measurement wavelength, in this case is relatively small.
Other potential sources of error include imperfect baffling of off-axis light, glow from electronic components such as LEDS on the telescope mount, or scattered light not produced by optical ghosts.
Due to the high signal-to-noise ratio of these measurements, these are expected to have minimal effect.

\section{Conclusion}
\label{sec:Conclusion}

We have designed and characterized an initial implementation of a collimated projector that can be used for the determination of the telescope instrumental response function. 
We have demonstrated success on both the Dark Energy Camera wide field imager on the 4 meter Blanco telescope and United States Naval Observatory Flagstaff Station's 1.3\,m reflector.
This design serves as the basis for the LSST collimated beam projector described in Ingraham et al \cite{InSt2016}.

In the future, we will need to demonstrate the ability to combine collimated beam projector data with dome flats.
This will require making measurements similar to the Dark Energy Camera data but at many pupil positions and at multiple wavelengths.
We also need to coordinate collimated beam projector measurements with on-the-sky calibration measurements.

\section{Acknowledgments}
MWC was supported by the National Science Foundation Graduate Research Fellowship
Program, under NSF grant number DGE 1144152. CWS is grateful to the DOE Office 
of Science for their support under award DE-SC0007881.
The authors would like to thank Dr. Anna Klales for a careful reading of an earlier version of this manuscript.

\bibliographystyle{spiebib}
\bibliography{references}

\end{document}